\documentclass[universe,article,accept,moreauthors,pdftex,10pt,a4paper]{mdpi}  

\usepackage[english]{babel}
\usepackage{t1enc}
\usepackage[utf8]{inputenc}
\usepackage[T1]{fontenc}
\usepackage{amsmath, amssymb, amsfonts}
\usepackage{indentfirst}
\usepackage{graphicx}
\usepackage{subfig}
\usepackage{color}
\usepackage{amsfonts}
\usepackage{array}
\usepackage{verbatim}
\newcolumntype{L}[1]{>{\raggedright\let\newline\\\arraybackslash\hspace{0pt}}m{#1}}
\newcolumntype{C}[1]{>{\centering\let\newline\\\arraybackslash\hspace{0pt}}m{#1}}
\newcolumntype{R}[1]{>{\raggedleft\let\newline\\\arraybackslash\hspace{0pt}}m{#1}}

\usepackage{anysize}

\newcommand{\nc}{\newcommand}
\nc{\td}{\mathrm{d}}
\nc{\kd}{\mathbf{k}}
\nc{\Kd}{\mathbf{K}}
\nc{\qd}{\mathbf{q}}
\nc{\Rd}{\mathbf{R}}
\nc{\rd}{\mathbf{r}}
\nc{\ud}{\mathbf{u}}
\nc{\vd}{\mathbf{v}}
\nc{\pd}{\mathbf{p}}
\nc{\Pd}{\mathbf{P}}
\nc{\Gd}{\mathbf{G}}
\nc{\Ad}{\mathbf{A}}
\nc{\Dd}{\mathbf{D}}
\nc{\Sd}{\mathbf{S}}
\nc{\md}{\mathbf{m}}
\nc{\Fd}{\mbox{\boldmath $F$}}
\nc{\Md}{\mbox{$\mathcal{M}$}}
\nc{\bd}{\mbox{\boldmath $\beta$}}
\nc{\Od}{\mbox{$\mathbf{\mathcal{O}}$}}
\nc{\od}{\mbox{\boldmath $\omega$}}
\nc{\Ddd}{\mbox{\boldmath $\underline{\underline{D}}$}}
\nc{\Mdd}{\mbox{\boldmath $\underline{\underline{M}}$}}
\nc{\Cd}{\mbox{\boldmath $\underline{\underline{C}}$}}
\nc{\Div}{\mathrm{div}}
\nc{\Rot}{\mathrm{rot}}
\nc{\Grad}{\mathrm{grad}}
\nc{\Det}{\mathrm{det}}

\let\originalleft\left 
\let\originalright\right 
\renewcommand{\left}{\mathopen{}\mathclose\bgroup\originalleft} 
\renewcommand{\right}{\aftergroup\egroup\originalright}

\firstpage{1}  
 
\articlenumber{x} 
\doinum{10.3390/------} 
\pubvolume{xx} 
\pubyear{2018} 
\copyrightyear{2018} 
\history{
	} 
 
\setitemize{parsep=6pt,itemsep=0pt,leftmargin=*,labelsep=5.5mm} 
\setenumerate{parsep=6pt,itemsep=0pt,leftmargin=*,labelsep=5.5mm} 
\setlist[description]{itemsep=0mm}

\pdfoutput=1 
\nolinenumbers 
\Title{ 
New exact solutions of relativistic hydrodynamics \\
for longitudinally expanding fireballs 
}
 
\Author{Tam\'as Cs\"org\H{o} $^{1,2,*}$, G\'abor Kasza $^{2}$, M\'at\'e Csan\'ad$^{3}$ and Zefang Jiang$^{4,5}$} 
 
\AuthorNames{Tam\'as Cs\"org\H{o}, G\'abor Kasza, M\'at\'e Csan\'ad and Zefang Jiang} 
 
\address{%
$^{1}$ \quad Wigner RCP, H-1525 Budapest 114, P.O.Box 49, Hungary\\ 
$^{2}$ \quad EKU KRC, H-3200 Gy\"ongy\"os, M\'atrai \'ut 36, Hungary\\ 
$^{3}$ \quad E\"otv\"os University, H-1117 Budapest, P\'azm\'any P. s. 1/A, Hungary\\ 
$^{4}$ \quad Key Laboratory of Quark and Lepton Physics, Ministry of Education, Wuhan, 430079, China\\ 
$^{5}$ \quad Institute of Particle Physics, Central China Normal University, Wuhan 430079, China 
} 
 
\corres{Correspondence: tcsorgo@cern.ch} 
 
\abstract{ 
We present new, exact, finite solutions of relativistic hydrodynamics for
longitudinally expanding fireballs for arbitrary constant value of the speed of
sound.  These new solutions generalize earlier, longitudinally finite, exact
solutions, from an unrealistic to a reasonable equation of state, characterized
by a temperature independent (average) value of the speed of sound.
Observables like the rapidity density and the pseudorapidity density are
evaluated analytically, resulting in simple and easy to fit formulae that can
be matched to  the high energy proton-proton and heavy ion collision data  at
RHIC and LHC.  In the longitudinally boost-invariant limit, these new solutions
approach the Hwa-Bjorken solution and the corresponding rapidity distributions
approach a rapidity plateaux.  
} 
 
 
\begin{document} 
 
 
 
 
\section{Introduction}

Some of the most renowned theoretical papers in high energy heavy ion physics
deal with exact solutions of perfect fluid hydrodynamics for a 1+1 dimensional,
longitudinally expanding fireball. 
In high energy collisions involving strong interactions, statistical
particle production rates were noted by Fermi already in 1950
~\cite{Fermi:1950jd}.
Landau, Khalatnikov and Belenkij predicted as early as 
in 1953-56~\cite{Landau:1953gs,Khalatnikov:1954,Belenkij:1956cd},
that in these collisions not only global but also local thermal equilibrium will
be a relevant concept and the related perfect fluid hydrodynamical modelling
will provide the framework for the future analysis of experimental data.

Landau's prediction about the perfect fluid behaviour in high energy heavy ion
collisions was, at that time quite unexpectedly,  fully confirmed by the
discoveries after the first four years of data-taking at Brookhaven National
Laboratory's Relativistic Heavy Ion Collider (RHIC) where the  picture of a
nearly perfect strongly interacting fluid  emerged from the first years of
observations of the four RHIC collaborations, BRAHMS ~\cite{Arsene:2004fa},
PHENIX ~\cite{Adcox:2004mh}, PHOBOS ~\cite{Back:2004je}, and STAR
~\cite{Adams:2005dq}.

When the ALICE Collaboration reported the first elliptic flow measurement at
the LHC in Pb+Pb collisions at $\sqrt{s_{NN}} = 2.76 $
TeV~\cite{Aamodt:2010pa}, they also noted the  similarity of the transverse
momentum dependence of the elliptic flow to earlier, lower energy data at RHIC
and noted its consistency with predictions of hydrodynamic models, confirming
the creation of a nearly perfect fluid of strongly interacting Quark Gluon
Plasma (sQGP) also at higher energies. More detailed measurements by ALICE,
ATLAS and CMS ~\cite{ALICE:2017jyt,Aad:2013fja,Chatrchyan:2013nka} gave further
support to the hydrodynamical paradigm in high energy heavy ion collisions at
RHIC and LHC. The domain of the validity of this paradigm is currently being
extended to describe proton-proton, as well as to proton/d/$^3$He-nucleus
collisions, from the highest LHC energies~\cite{ALICE:2017jyt} down to
the lowest RHIC energies ~\cite{Adare:2013piz,Aidala:2017pup,Aidala:2017ajz}.

The status of the applications of hydrodynamical modelling
to high energy collisions was reviewed recently ~\cite{deSouza:2015ena},
including  a review of exact solutions as well, but 
focussing mostly on the numerical solutions of relativistic hydrodynamics, as
well as on the conceptual and open questions.

To define the context of our new exact solutions described in the body of the
manuscript, let us dedicate the next section to a brief, but specialized
overview of the status of the search for exact solutions of fireball
hydrodynamics and the application of these results to high energy hadron-proton
and heavy ion collisions. The equations of perfect fluid hydrodynamics are
summarized in the subsequent section.  Next, we present our new exact,
accelerating family of solutions of relativistic hydrodynamics, pointing out
their finite range in rapidity,
but without detailing their derivation.
 Then we evaluate the observable rapidity and
pseudorapidity densities near midrapidity, discuss and compare the results with
experimental data on proton-proton collisions at the $\sqrt{s} = $$7$ and $8$
TeV at LHC. Finally we summarize and conclude.

\section{A brief overview of the status of the field} 
\label{s:overview}

\subsection{Exact solutions with boost-invariant flow profiles}
Hydrodynamics typically predicts strong anisotropies in the angular
distribution of produced particles, with a predominantly longitudinal
expansion, giving a motivation to study exact solutions of perfect fluid
hydrodynamics in 1+1 dimensions, characterized by a temporal and a spatial
dimension, where the spatial dimension corresponds to the main axis of the
expansion, usually taken as the longitudinal direction.

The ultra-relativistic limit of the hydrodynamical models was, as far as we
know,  solved for the first time by R. C. Hwa as early as in 1974. Hwa noted
that in the limit of infinite colliding energies, the resulting rapidity
distribution of produced hadrons will be flat,  a constant value, also called the
rapidity plateaux~\cite{Hwa:1974gn}.  Independently, but several years later,
Bjorken rediscovered the same boost-invariant  asymptotic
solution~\cite{Bjorken:1982qr}.   In addition, rather importantly Bjorken also
realized that the rapidity density can be used to estimate the initial energy
density in these reactions~\cite{Bjorken:1982qr}.  By now, the resulting
Bjorken energy density estimate became the top-cited phenomenological
formula in high energy heavy ion physics.

The Hwa-Bjorken boost-invariant solution was extended to include
transverse expansion dynamics as well. Certain exact solutions were obtained
by Bír\'o that considered the case of transverse expansion on top of
a longitudinally boost invariant flow during a time period with constant pressure.
Such a scenario was thought to be relevant
during the course of a first order QGP --  hadron gas phase
transition~\cite{Biro:2000nj}.  Gubser found a similar,
longitudinally boost-invariant but transversally expanding solution
for a massless relativistic gas, characterized by the equation of
state $\varepsilon = 3 p$, where $\varepsilon$ denotes the energy
density and $p$ stands for the pressure.  This choice of the equation
of state corresponds to conformally symmetric solutions
~\cite{Gubser:2010ze,Gubser:2010ui},  where the transverse flow is
initially vanishing, and it is increasing at smaller transverse
distances nearly linearly, that however it does not keep on increasing but
changes to decreasing to zero at large distances.  Solutions for very
viscous plasmas (cold plasma limit) were also found that recover the
perfect fluid solution of Gubser in the limit of vanishing viscosity
~\cite{Marrochio:2013wla}.

Using the equation of state of a relativistic ideal gas and conformal
symmetry, analytic axially symmetric solutions with non-zero
vorticity was described in ref.  ~\cite{Hatta:2014gqa} and detailed
in ref.  ~\cite{Hatta:2014gga} , with the aim that the
phenomenological extension of these solutions to longitudinally
boost-invariant but elliptically expanding case may give clues to
understand vorticity effects and the elliptic flow data in heavy ion
collisions at RHIC and at LHC. Results from this approach were 
reviewed briefly in ref. ~\cite{Hatta:2016czn}.

A new analytic solution of relativistic hydrodynamics for perfect
fluids with conformal symmetry and equation of state for a massless
relativistic gas, $\varepsilon = 3 p$ was presented very
recently~\cite{Bantilan:2018vjv}.  This new solution corresponds to
rotations  of the fluid around two different vortices (axis),
generalizing simultaneously the non-rotating  solutions of refs.
~\cite{Gubser:2010ze,Gubser:2010ui}, as well as the  rotating
solutions described in  ~\cite{Nagy:2009eq} and more recently
described and detailed in ref.~\cite{Hatta:2014gqa,Hatta:2014gga}.
The possibility of describing relativistic fluids with one and two
vortices prompts the question, if perhaps generalized exact formulae
can be derived that describe expanding and locally rotating
relativistic fluids, that may have more than two vortices?

In this context, it is also of importance to mention that some of the exact
solutions include solutions with the lattice QCD equation of state, using
actually an even broader class of equations of state, where the speed of sound
as well as the constant of proportionality between the pressure and the energy
density may be an arbitrary but temperature dependent function.  As far as we
know, the first solution with a temperature dependent speed of sound was given
in ~\cite{Csorgo:2001xm} , using the equation of state $\varepsilon = \kappa(T)
p$ for a system that had conserved charge density $n$ and the pressure was
given as $p = n T$.  Independently, this kind of equation of state was
generalized for the  $\mu_B = 0$ case, when only the entropy density $\sigma$
is locally conserved, and the pressure is proportional to the entropy density, $p \propto
\sigma T$ with $\varepsilon = \kappa(T) p$ in ref.  ~\cite{Csanad:2012hr},
opening the way to use lattice QCD equation of state in several subsequent
exact solutions of relativistic as well as non-relativistic hydrodynamics,
~\cite{Csorgo:2015scx,Nagy:2016uiz,Csorgo:2018tsu}.

Actually, the non-vanishing total angular momentum of the fireball created in
high energy heavy ion collisions is an interesting problem in itself.  It is
clear that infinite, longitudinally boost-invariant sources cannot have an
angular momentum perpendicular to the impact parameter plane, because their
moment of inertia in the beam direction is infinite.  So the search for
rotating solutions is indirectly coupled to the search for 
non-boost-invariant, finite exact solutions of fireball hydrodynamics.

The first exact solution of relativistic hydrodynamics was found by Nagy,
when searching for simultaneous solutions of relativistic hydrodynamics
and the relativistic collisionless Boltzmann equation~\cite{Nagy:2009eq}.
This searching method generalized the non-relativistic search method
of Csizmadia, Cs\"org\H{o} and Luk\'acs~\cite{Csizmadia:1998ef}
to the relativistic kinematic domain, with surprising success.

\subsection{Recent results on rotating, non-relativistic solutions}
Several exact solutions of rotating, non-relativistic finite fireballs were
obtained recently, both for spheroidal and for ellipsoidal symmetries,
corresponding to expanding and rotating spheroidal ~\cite{Csorgo:2015scx} or
triaxial, ellipsoidal density profiles ~\cite{Nagy:2016uiz}, that all go back
to the spherically symmetric, non-rotating and collisionless solution of
fireball hydrodynamics of ref.~\cite{Csizmadia:1998ef}.  These solutions are
finite and aspire to explain some of the scaling properties of the single
particle slope parameters~\cite{Csizmadia:1998ef,Csorgo:2018tsu}, 
the elliptic flow and the
Hanbury Brown-Twiss (HBT) radii. 

Recently new exact solutions including  a realistic, lattice QCD based equation
of state and the effects of angular momentum ~\cite{Csorgo:2016ypf} were found
to describe well the mass systematics of single particle spectra from a
strongly interacting quark-gluon plasma being converted to a multi-component
hadron gas ~\cite{Csorgo:2018tsu} .  It is interesting to note that although
these non-relativistic exact solutions of rotating fireball hydrodynamics were
found very recently, they were obtained using analogies with relativistic exact
solutions of rotating fireball hydrodynamics, that were found first by Nagy in
2009, when looking for simultaneous solutions of relativistic hydrodynamics and
the relativistic collisionless Boltzmann equation ~\cite{Nagy:2009eq}. 

As the medium cools down, quark and gluon degrees of freedom get confined to
hadrons.  After such a re-confinement, soon a  hadrochemical freeze-out may be
reached, where the equations describing local energy-momentum and entropy
conservation are supplemented with the continuity equation for each of the
locally conserved particle densities.  This problem was considered recently in
refs.  ~\cite{Csorgo:2016ypf,Csorgo:2018tsu}, both in the relativistic and in
the non-relativistic kinematic case, successfully explaining the
mass-dependence of the slope parameters of single-particle spectra from an
exact solution of non-relativistic hydrodynamics, showing also that many
features of these solutions are similar to the system of equations of the
relativistic fluid dynamics. 

One of the important observations is that at late time, these non-relativistic
solutions approach an asymptotic, 3-dimensional Hubble or rotating Hubble flow
profiles, which are solutions of hydrodynamics in the relativistic kinematic
domain as well. In the longitudinal direction, the Hubble flow profile and the
Hwa-Bjorken scaling flow profiles coincide. 
One of the important limitations of the boost-invariant Hwa-Bjorken solution
~\cite{Hwa:1974gn,Bjorken:1982qr}   is its infinite range along the axis of
expansion.  In contrast, the non-relativistic, spherically symmetric
Zim\'anyi-Bondorf-Garpman solution ~\cite{Bondorf:1978kz} is similarly simple,
but it is a compact and finite,  spherically symmetric solution.  Actually, the
velocity field is a linear, Hubble-type flow field with $\mathbf{ v} \propto \mathbf{r}$ in both
of the relativistic Hwa-Bjorken and in the non-relativistic
Zim\'anyi-Bondorf-Garpman
solution~\cite{Hwa:1974gn,Bjorken:1982qr,Bondorf:1978kz}.  
Thus, it would be worthwhile to search for the relativistic generalizations
of the non-relativistic but rotating exact solutions given in refs.
~\cite{Csorgo:2015scx,Nagy:2016uiz,Csorgo:2016ypf,Csorgo:2018tsu}.

\subsection{Relativistic solutions without longitudinal boost-invariance and rotation }

One of the natural themes when searching for new exact solutions
of relativistic hydrodynamics was to generalize the Hwa-Bjorken 
boost-invariant longitudinal
flow to a finite, realistic, accelerating longitudinal flow profile. 
Initially, such solutions were found by breaking the boost-invariance
of the temperature or the (entropy)density distributions, while
keeping the pressure and the flow-field 
boost-invariant~\cite{Csorgo:2002ki,Csorgo:2003rt,Csorgo:2003ry}.

In these solutions, the Hubble flow field  remained an important
element, but one obtained generalized, spatially inhomogeneous
temperature and density profiles in new, self-similar, $1$$+$$1$
dimensional and $1$$+$$3$  dimensional axially
symmetric solutions by Cs\"org\H{o}, Grassi, Hama and Kodama in refs~\cite{Csorgo:2002ki,Csorgo:2003rt}, or to $1$$+$$3$ dimensional
ellipsoidally symmetric solutions by Cs\"org\H{o}, Csernai, Hama and
Kodama of ref.  ~\cite{Csorgo:2003ry}. Later, we shall refer to these
as the CGHK and the CCHK solutions,  respectively.
In these solutions, the equation of state was reasonable as well, as
the constant of proportionality between the pressure and the energy
density had arbitrary value.  The transverse momentum spectra of
identified pions, kaons and protons, as well as elliptic flow data
and HBT radii, the characterisic length-scales measured by
Bose-Einstein correlation functions were shown to be well described
by the CCHK solution~\cite{Csorgo:2003ry}, as demonstrated in
~\cite{Csanad:2009wc}. Even the direct photon spectra and the
elliptic flow of direct photons was shown to be well described by
this CCHK solution ~\cite{Csanad:2011jq}.
However, from the theoretical point of view, the
main limitation of the CCHK solution was its non-accelerating
property, due to the presence  the Hubble flow field, that can be
seen as an asymptotic attractor for late stages of fireball
explosions.

Let us note here that hydrodynamics was successfully applied to describe the
double-differential rapidity and transverse mass spectra in hadron+proton
collisions at the surprisingly low energy of $\sqrt{s}\simeq 22$ GeV already in
1997, using the Buda-Lund hydro model ~\cite{Csorgo:1995bi}, 
that also predicted a special coupling
between longitudinal dynamics and transverse expansion.  The hydrodynamically
predicted decrease of the slope of the transverse momentum spectra at forward
and backward rapidities was experimentally observed ~\cite{Agababyan:1997wd}. 
Although phenomenologically very successful, this
Buda-Lund hydrodynamical model used a boost-invariant longitudinal flow
profile, but a non-boost-invariant longitudinal chemical potential or density.
Subsequently, the time evolution, 
corresponding to the Buda-Lund hydro model of ref.  ~\cite{Csorgo:1995bi} 
was related to the time evolution in the  CCHK exact solution ~\cite{Csorgo:2003ry}. 

\subsubsection{\it Landau hydrodynamics}
One of the research directions to describe the finite rapidity or
pseudorapidity distributions was due to Wong and collaborators, who first
improved the evaluation of the rapidity density ~\cite{Wong:2008ex} and
subsequently studied the matching of the inwards-moving shock-wave of
Khalatnikov's solution with the outward-moving regular solution used by Landau
to obtain an approximately Gaussian rapidity distribution, using the massless
relativistic ideal gas equation of state, $\varepsilon = 3 p$
~\cite{Wong:2014sda}.

By using the revised Landau hydrodynamic model and taking into
account the effect of leading particles as independent Gaussian sources
at forward and backward rapidities, ref.  ~\cite{Jiang:2013tra}
explained the  pseudorapidity distributions of produced charged particles
at different centralities in $\sqrt{s_{NN}}= 200$ GeV Cu+Cu and Au+Au collisions
at RHIC. Similarly good descriptions of the pseudorapidity distributions
were obtained at lower colliding energies, $\sqrt{s_{NN}} = 130 $ and 62 GeV
~\cite{Jiang:2013vda}. 

Starting from about 2014, the Landau hydrodynamics was revisited and tested
successfully on proton-proton collisions  as well, from $\sqrt{s} = $ 23 to 900
GeV, ~\cite{Jiang:2014sta,Jiang:2014wza}. The low energy limit of these good
fits also moved further down  in the case of heavy ion collisions to
$\sqrt{s_{NN}} = $ 19.6  and 22.4 GeV Au+Au and Cu+Cu
collisions~\cite{Jiang:2014uya}, and the range also was extended upwards to
$\sqrt{s_{NN}} = $ 2.76 TeV Pb+Pb collisions at CERN LHC. In order to achieve
a successful fit, a
Gaussian contribution explained as leading particle contribution had to be
added to the Landau hydrodynamical calculations both at forward and at backwards
rapidities ~\cite{Hai-Li:2014opa}.  Landau hydrodynamics was generalized also
for an arbitrary constant speed  of sound and it was shown to give a similarly
reasonable description of the pseudorapidity densities at 2.76 TeV Pb+Pb
collisions ~\cite{Jiang:2014mqw}.

\subsubsection{\it Results from the BJP solution}
Another accelerating, one-parameter family of analytic solutions
interpolating between the boost-invariant Bjorken picture and the non
boost-invariant one, similar to the Landau profile was described by
Bialas, Janik and Peschanski in ~\cite{Bialas:2007iu}, referred to as
the BJP solution.  Using the Khalatnikov potential, 
an analytic formula was derived 
from the BJP solution for the rapidity dependence of
the entropy density, $dS/dy$~\cite{Beuf:2008vd}.  

Recently, it was found that the BJP solution describes the
longitudinal evolution in relativistic heavy ion collisions both at
RHIC and at LHC, when a leading particle effect is included, for a
realistic, constant value of the speed of sound, including the
centrality dependence in Cu+Cu and  Au+Au collisions at $\sqrt{s_{NN}} =
200 $ GeV and in Pb+Pb collisions at $\sqrt{s_{NN}} = 2.76 $
TeV~\cite{Jiang:2015kba}. The same BJP solutions were shown to
describe the longitudinal evolution and the resulting pseudorapidity
distributions in proton-proton collisions as well, from 23 GeV to 7
TeV ~\cite{Jiang:2015rca}, when the leading particle effect was
corrected for in the forwards and backwards direction.  Thus the BJP
solution of relativistic hydrodynamics describes well the
longitudinal observables both in proton-proton and in heavy ion
collisions, from the lowest to the highest presently available
energies~\cite{Jiang:2016gan,Jiang:2016mbx}.

The elliptic and higher order flows were also obtained analytically
from the BJP solution, using the assumption that the entropy is
transversally conserved.  This assumption holds for the asymptotic
Hwa-Bjorken flows but, as far as we can see, it is not a generally
valid property of accelerating longitudinal flows, as it neglects the
important correction due to work done by central fluid elements on
the surface~\cite{Peschanski:2009tg}. Despite these successes of the BJP
solution, these solutions have not yet been connected to possible
estimations of the initial energy density of high energy proton-proton
or heavy ion collisions.

\subsubsection{\it Results from the CNC solution}

Motivated by  searching for solutions with finite rapidity distributions and
corrections to the initial energy density estimate of Bjorken, an exact and
explicit, longitudinally accelerating solutions of relativistic hydrodynamics,
a one-parameter family of analytic solutions was presented for $1$$+$$1$
dimensional explosions by Cs\"org\H{o}, Nagy and Csan\'ad in
ref.~\cite{Csorgo:2006ax} and detailed in ref.~\cite{Nagy:2007xn}, referred to
as the CNC solution.  This CNC solution  generalized the Bjorken flow field of
$v_z = \tanh(\eta_x) = r_z/t$ to $v_z = \tanh(\lambda \eta_x)$, where $v_z$  is
the longitudinal component of the four-velocity field, $\eta_x = (1/2)
\ln\left[(t+r_z)/(t-r_z)\right]$ is the space-time rapidity and $\lambda$ is a
parameter that characterizes the acceleration of the fluid. This one parameter
family of solutions had five different domains of applicability, where
different domains were characterized by the value of the acceleration parameter
$\lambda$, the number of spatial dimensions $d$ and the parameter of the
equation of state, $\kappa$.

Almost simultaneously,  Borshch and Zhdanov published exact solutions using the
superhard $\kappa = 1$ equation of state, that included the CNC solution
~\cite{Csorgo:2006ax,Nagy:2007xn} 
in certain limiting cases, and some other more general solutions also where the
equation of state was more realistic~\cite{Borshch:2007uf}, but the
phenomenology related to observables in high energy heavy ion collisions has
not yet been developed as far as we know.  It would be interesting to see how
these Borshch-Zhdanov or BZ solutions compare to experimental data at RHIC and at LHC.

In one out of these five classes of the CNC solutions~\cite{Csorgo:2006ax}, the
parameter $\lambda$ was arbitrary, its value could be determined from fits to
experimental data. This CNC solution not only resulted in a realistic and
straightforwardly usable result for both the rapidity and the pseudorapidity
densities, but these results were also applied to get important and large
corrections (factors of 5-10) to the initial energy density estimate of
Bjorken. These corrections take into account the work done by the central fluid
elements  on the surface.  Unfortunately, these CNC solutions have also a major
shortcoming, namely that the acceleration parameter $\lambda$ became a free fit
parameter only for the superhard equation of state of  $\kappa = 1$,
$\varepsilon = p$. In this case, the speed of sound is equal to the speed of
light, so the investigation was thought to be rather academic. Nevertheless,
the equation of state dependence of the initial energy density estimate was
exactly determined for $\kappa = 1$ case,  and it was determined to yield
important corrections to Bjorken's energy density estimate, a factor of 10 at
RHIC energies. However, the dependence of this correction factor on
the equation of state has not yet been determined exactly, 
this factor was only conjectured so far, but its numerical value 
was found to yield large corrections, of the order of
15~\cite{Csorgo:2008pe,Csanad:2016add}. In this work, we present a solution
that may be the basis of an exact derivation of an equation of state and acceleration
dependent correction to Bjorken's initial energy density estimate,
as one of the motivations of our study was 
to rigorously derive the equation of state dependence of the corrections to
Bjorken's energy density estimate.  So the search for accelerating solutions
has been continued, and some of the new results that have been achieved
so far are presented in the body of this manuscript.

Recently, the formulas obtained by the CNC solution, that include the
work effect but have an unrealistic, superhard equation of state,
were also shown to describe surprisingly well the pseudorapidity
densities not only in proton-proton collisions at the LHC energies
~\cite{Csanad:2016add,Ze-Fang:2017ppe}. At the same time, the
pseudorapidity densities of heavy ion collisions were described
as well, including RHIC and LHC
energies as well, from $\sqrt{s_{NN}} = 130 $ GeV to $\sqrt{s} = 8 $
TeV. These results may indicate, that the longitudinal expansion dynamics of high
energy proton-proton and heavy ion collisions is surprisingly
similar. In addition, they indicate that the CNC solution works
surprisingly well, much better than expected when compared to the
experimental data, so in some sense its final stage at the freeze-out
is likely very similar to a final state obtained from solutions of
relativistic hydrodynamics with more realistic equations of state.
So the success of the CNC fits of ref.  ~\cite{Ze-Fang:2017ppe}
motivated our current paper to search for accelerating, finite, exact
solutions of $1$$+$$1$ dimensional relativistic hydrodynamics with
reasonable (but yet temperature independent) value for the speed of
sound.

Motivated by the success of Landau hydrodynamics
~\cite{Wong:2014sda,Jiang:2013tra} and pointing out the surprising
and almost unreasonable success of the CNC solution
~\cite{Csorgo:2006ax,Nagy:2007xn,Csorgo:2008pe} to describe the
pseudorapidity distributions from proton-proton to heavy ion
solutions ~\cite{Ze-Fang:2017ppe,Csanad:2016add}, even with a
superhard equation of state, $\varepsilon = p$, we suspected that the
CNC solution may be very close to certain analytic solutions that
have arbitrary equation of state but the fluid rapidity $\Omega $ is
still nearly proportional to the coordinate-space rapidity $\eta_x$.
Indeed, we found that such exact solutions exist and they are 
surprisingly close in shape to the
CNC solutions, given the approximate proportionality between the
fluid rapidity and the space-time rapidity in these solutions, if we
determine these not too far away from mid-rapidity.

\section{Equations of relativistic hydrodynamics}
\label{s:hydro-CKCJ} 
 
For perfect fluids, relativistic hydrodynamics expresses the local conservation of energy, momentum and entropy :
\begin{align} 
        \partial_{\nu}T^{\mu \nu}&=0, \label{eq:energy-momentum-conservation} \\
        \partial_{\mu}\left(\sigma u^{\mu}\right)&=0, \label{eq:continuity-entropy-density}
\end{align} 
where $T^{\mu \nu}$ is the energy-momentum four-tensor,  $\sigma = \sigma(x)$
is the entropy density and  $u^{\mu}$ stands for the four-velocity normalized
as $u^{\mu}u_{\mu}=1$.  These fields depend on the four-coordinate
$x^{\mu}=(t,\mathbf{r}) = \left(t,r_x,r_y,r_z\right)$. The
four-momentum is denoted by $p^{\mu} = (E_p, \mathbf{p}) =
\left(E_p,p_x,p_y,p_z\right)$  with an on-shell energy $E_p = \sqrt{m^2 +
\mathbf{p}^2}$, where $m$ is the mass of a given type of observable particle.
Eq.~\eqref{eq:energy-momentum-conservation} stands for
local conservation of  energy and momentum, while 
eq.~\eqref{eq:continuity-entropy-density} expresses the local conservation of
entropy density, and the lack of dissipative terms in perfect fluid
hydrodynamics.
 
The energy-momentum four-tensor of perfect fluids is denoted by  
$T^{\mu\nu}$ which reads as follows:  
\begin{equation} 
T^{\mu \nu}=\left(\varepsilon+p\right)u^{\mu}u^{\nu} - pg^{\mu \nu}, 
\end{equation} 
where the metric tensor is denoted by $g^{\mu 
\nu}=\textnormal{diag}(1,-1,-1,-1)$, and as already explained in 
section ~\ref{s:overview}, 
the energy density is denoted by $\varepsilon$ and $p$ stands for the pressure.
 
The basic equations of relativistic, perfect fluid hydrodynamics  is
frequently applied also in the case when there are (perhaps several)
locally conserved charges.  For example, after hadrochemical
freeze-out at the temperature $T_{chem}$, the particle number density
$n_i(x)$ is locally conserved for each of the frozen-out hadronic
species indexed by $i$. In this case, the entropy equation is
supplemented by the local charge conservation equations, expressed as
follows: 
\begin{equation} 
\partial_{\mu}\left(n_i u^{\mu}\right)=0, \qquad \mbox{\rm for} \quad {T\le T_{chem},} \quad {i=1,2,...,j} 
\label{e:cont-i} 
\end{equation} 
where $n_i$ is the particle density of the $i^{th}$ hadron, and $j$
counts that how many kind of hadrons are frozen out hadrochemically.  
For such a mixture of hadrons, $m_i$ denotes the mass for hadron type $i$.

The equation expressing the local conservation of energy and momentum 
can be projected to a component parallel to $u^{\mu}$ that yields the energy equation: 
\begin{equation} 
\partial_{\mu}\left(\varepsilon u^{\mu}\right) + p \partial_{\mu}u^{\mu} = 0,  \label{e:rel-energy} 
\end{equation}\ 
while the component pseudo-orthogonal to the four-velocity field 
yields the relativistic Euler equation: 
\begin{equation} 
\left(\varepsilon + p\right) u^{\nu }\partial_{\nu}u^{\mu}=\left(g^{\mu \nu} - u^{\mu}u^{\nu} \right)\partial_{\nu}p. 
\label{e:rel-Euler} 
\end{equation}

We have five differential equations for six independent quantities,
the three spatial components of the velocity field and the entropy
density, the energy density and the pressure, 
($v_x(x)$, $v_y(x)$, $v_z(x)$,$\sigma(x)$, $\varepsilon(x)$, $p(x)$). 
Thus the system of partial differential equations that corresponds to
relativistic hydrodynamics is closed by 
the equation of state (EoS), that  defines the connection between two of the 
six unknown fields. 
Throughout this paper, let us assume that the energy density is proportional to the pressure:
\begin{equation}
	\varepsilon=\kappa p,
\end{equation}
where $\kappa$ is assumed to be a temperature independent constant. 
Thus in this work we search for $1$$+$$1$ dimensional exact solutions
of relativistic hydrodynamics that correspond to a generalized,
$\varepsilon = \kappa p$ equation of state, where the realistic value of
the speed of sound is about $c_s = 1/\sqrt{\kappa} = 0.35\pm
0.05$~\cite{Adare:2006ti}, so the reasonable range of $\kappa$ is
approximately from 6 to 11.
This EoS closes the system of partial differential equations
that constitute relativistic hydrodynamics.

Although in a fully realistic solution, we should use the lattice QCD
equation of state, with the speed of sound being a temperature
dependent function, similarly to refs.
~\cite{Csorgo:2001xm,Csanad:2012hr,Csorgo:2015scx,Nagy:2016uiz,Csorgo:2016ypf,Csorgo:2018tsu}.
We postpone the analysis of such shockwaves to for later, more detailed 
investigations.  The success of the BJP
~\cite{Jiang:2015kba,Jiang:2015rca,Jiang:2016gan} and the CNC
solutions ~\cite{Ze-Fang:2017ppe,Csanad:2016add} in describing
pseudorapidity densities using a temperature independent constant of
proportionality $\kappa$ between the energy density and the pressure
provides further, independent support for these investigations.

With the help of the  fundamental equation of thermodynamics,  
a new variable, the temperature $T$ is introduced as
\begin{equation}
	\varepsilon + p = \sigma T + \sum_{i=1}^j \mu_i n_i .\label{e:thermo}
\end{equation}
In what follows, we do not consider the effects of conserved charges and we
assume that the corresponding chemical potentials vanish, $\mu_i = 0$.

Let us note here that we intended to search for self-similar solutions,
and in the next section we actually describe a rich family of new
and exact solutions, that have a trivial proper-time dependence but
also that depend on the coordinates predominantly through a certain scaling variable
$s$. At the end we will describe solutions that break self-similarity,
nevertheless they do it with explicit factor that vanishes around mid-rapidity, 
so the concept of self-similarity and its actual breaking 
will make the investigation of the question of self-similarity of the solution
an interesting one.

The definition of the scaling variable $s$ is that it
has a vanishing co-moving derivative:
\begin{equation}
u_{\mu}\partial^{\mu}s=0.
\end{equation}
If $s$ is a scaling variable, that satisfies the above equation then obviously
any $s^\prime=s^\prime(s)$ that is a function of $s$ only may also be introduced as a new scaling variable,
given that 
\begin{equation}
	u_{\mu}\partial^{\mu}s^\prime= \partial_s s^\prime
	u_{\mu}\partial^{\mu}s = 0.
\end{equation}

In $1$$+$$1$ dimension, let us rewrite the equations of relativistic hydrodynamics
with the help of the Rindler coordinates $(\tau,\eta_x)$ for the temperature $T$
and for the rapidity of the fluid $\Omega$.
The longitudinal proper-time, referred to in what follows simply as the proper-time ($\tau$),
the coordinate-space rapidity $\eta_x$ and the fluid rapidity $\Omega$ are defined as 
\begin{align}
	\tau&=\sqrt{t^2-r_z^2},\\
	\eta_x&=\frac{1}{2}\textnormal{ln}\left(\frac{t+r_z}{t-r_z}\right),\\
	\Omega &=\frac{1}{2}\textnormal{ln}\left(\frac{1+v_z}{1-v_z}\right).
\end{align}
The fluid rapidity $\Omega$ relates to the four-velocity and to the three-velocity as
\begin{eqnarray}
	u^{\mu} &=& 
	\left(\textnormal{cosh}\left(\Omega\right),\textnormal{sinh}\left(\Omega\right)\right), \\
	v_z & = &\textnormal{tanh}\left(\Omega\right). 
\end{eqnarray}
For notational clarity, let us also highlight here the definition of the observables, 
that are determined by measuring particle tracks in momentum-space. The pseudorapidity $\eta_p$ and the rapidity $y$
of a final state particle with mass $m$ and four-momentum $p^\mu$ are defined as
\begin{align}
	\eta_p & = \frac{1}{2}\textnormal{ln}\left(\frac{p +p_z}{p-p_z}\right),\\
	y & = \frac{1}{2}\textnormal{ln}\left(\frac{E +p_z}{E-p_z}\right),
\end{align}
where the modulus of the three-momentum is denoted by $p = |\mathbf{p}| = \sqrt{p_x^2 + p_y^2 + p_z^2}$.

In the temperature independent speed of sound or $\kappa(T) = const $ approximation, the relation between the
pressure, temperature and entropy density simplifies as follows:
\begin{equation}
p=\frac{T\sigma}{1+\kappa}.
\end{equation}

From now, let us assume that the fluid rapidity 
depends only on the coordinate space rapidity, 
but not on the proper time, so that  
\begin{eqnarray}
	\Omega & = & \Omega(\eta_x),
\end{eqnarray}
With the help of the above assumption, the relativistic energy and Euler equations
can be rewritten for the temperature and for the fluid rapidity in terms of the Rindler coordinates,
using the notation $T = T(\tau,\eta_x)$ and $\Omega=\Omega(\eta_x)$. 
\begin{eqnarray}
	\label{eq:energy-temperature}
	\partial_{\eta_x}\Omega + \kappa \left(\tau \partial_{\tau} + 
	\tanh\left(\Omega-\eta_x\right)\partial_{\eta_x}\right)\ln\left(T\right) & = &0,\\
	\label{eq:Euler-temperature}
	\partial_{\eta_x}\ln\left(T\right)+\tanh\left(\Omega-\eta_x\right)\left(\tau\partial_{\tau}\ln\left(T\right) 
	+ \partial_{\eta_x}\Omega\right)& = &0.
\end{eqnarray}
In this way we obtained a set of partial differential equations for
the temperature and for the fluid-rapidity using the variables
$(\tau,\eta_x)$.  The solutions of these hydrodynamical equations
are presented in the next section.

Let us clarify, that we do not present the details of the derivation
of these solutions in this manuscript: these complicated details go 
well beyond the scope of the presentation of the results. 
However, the validity of these solutions is straigthforward to check
by substituting them to the temperature and the Euler equations of
Eqs.~  (\ref{eq:energy-temperature},\ref{eq:Euler-temperature}) and to the
entropy equation of Eq.~(\ref{eq:continuity-entropy-density}).
Note also that the above temperature and Euler equations,
Eqs.~  (\ref{eq:energy-temperature},\ref{eq:Euler-temperature}) were derived 
in Rindler coordinates before, in arbitrary $d$ number dimensions,
as given in Appendix A of ref. ~\cite{Nagy:2007xn}.

\section{A new family of  exact solutions 
of relativistic hydrodynamics}

We have discovered the following family of exact and explicit solutions of the equations of relativistic hydrodynamics:
\begin{eqnarray}
	\eta_x(H)  & = & \Omega(H) -H, \\ 
		\label{e:etaH}
	\Omega(H)  & = & \frac{\lambda}{\sqrt{\lambda-1}\sqrt{\kappa-\lambda}}
	\textnormal{arctan}\left(\sqrt{\frac{\kappa-\lambda}{\lambda-1}}\textnormal{tanh}\left(H\right)\right), \\ \label{e:OmegaH}
	\sigma(\tau,H)&= & 
	\sigma_0 \left(\frac{\tau_0}{\tau}\right)^{\lambda}\mathcal{V}_{\sigma}(s)
		\left[1+\frac{\kappa-1}{\lambda-1}\textnormal{sinh}^2(H)\right]^{-\frac{\lambda}{2}},\\	\label{e:sigmasol}
	T(\tau,H)  & = & T_0 
	\left(\frac{\tau_0}{\tau}\right)^{\frac{\lambda}{\kappa}} 
	\mathcal{T}(s)
		\left[1+\frac{\kappa-1}{\lambda-1}\textnormal{sinh}^2(H)\right]^{-\frac{\lambda}{2\kappa}},\\ \label{e:Tsol}
	\mathcal{T}(s) & = & \frac{1}{\mathcal{V}_{\sigma}(s)},\\ \label{e:scalingsol}
	s(\tau,H) & = & \left(\frac{\tau_0}{\tau}\right)^{\lambda-1} 
		\textnormal{sinh}(H)\left[1 + \frac{\kappa-1}{\lambda-1}\textnormal{sinh}^2(H)\right]^{-\lambda/2}.
		\label{e:sH}
\end{eqnarray}
A new property of the above equations is that the space-time rapidity $\eta_x$ dependence of the hydrodynamic fields
are given as parametric curves with explicit proper-time dependence. If one intends to analyse these solutions
at a given proper-time $\tau$, one may plot for example the $(\eta_x(H),F(\tau,H))$ 
parametric curve,
where $F$ stands for any of the hydrodynamic fields: $F =\left\{ T, \sigma, p, \varepsilon, \Omega, v_z, ...\right\}$.
The parameter of these curves, 
$H = \Omega - \eta_x$ has a clear physical meaning, it is the difference of the fluid rapidity
$\Omega$ and the space-time rapidity $\eta_x$. Some special properties of these parametric curves is analyzed in
the next subsection.

The quantity $\lambda $ is a constant of integration, that is a measure of the relativistic acceleration, and as we shall see
subsequently, this parameter also controls the width of the rapidity distribution and it can be determined from the fits to
experimental data. It is worth to note here that these solutions are obtained in the physical region, where $1 \le \lambda < \kappa$,
given that experimental data indicate $\lambda \simeq 1.05- 1.2 $ at RHIC and LHC energies, while in 200 GeV Au+Au collisions,
the average value of the speed of sound was measured by the PHENIX collaboration, yielding $\kappa = 1/c_s^2 \approx 10$~\cite{Adare:2006ti}.
The constants of normalization, $\sigma_0$ and $T_0$  are chosen to denote the initial conditions, the value 
of the entropy density and temperature at the initial proper-time $\tau = \tau_0$ and at mid-rapidity, where $\eta_x = \Omega = H = 0$. 
The scaling functions for the entropy and the temperature profile
are denoted as $\mathcal{V}_{\sigma}(s)$ and $\mathcal{T}(s)$, but as indicated in the above solution, only one of them
(for example the temperature profile function) can be chosen arbitrarily. The scaling variable $s$ above is normalized 
so that its proper-time dependent prefactor is unity at the initial proper-time, $\tau = \tau_0$ and its value vanishes
at midrapidity, $s(\eta = 0) = 0 $ as usual.

\subsection{Discussion and limiting cases}

One of the most important properties of the above equations is that they 
describe a family of exact solutions: for every positive definite, univariate
scaling function of the temperature $\mathcal{T}(s)$ one finds a corresponding
solution, so that (apart from an overall normalization factor) the temperature
and the entropy density depends predominantly on the space-time rapidity $\eta_x$ only
through this scaling function $\mathcal{T}(s)$. This property is inherited from
the 1+1 dimensional family of solutions of relativistic hydrodynamics described
by Cs\"org\H{o}, Grassi, Hama and Kodama in refs.
~\cite{Csorgo:2002ki,Csorgo:2003ry},  that are recovered in the 
$H \ll 1$ and $\lambda \rightarrow 1 $
limit of our new solutions. However, in our case, additional factors are also present
in the solutions, that break their self-similarity explicitely. 

The space-time rapidity dependence of our new solutions is given by Eqs.~(\ref{e:sH}-\ref{e:etaH}) .
These equations describe parametric curves. This point requires a careful analysis as the physical solutions
are to be given not as parametric curves, but as functions of the space-time rapidity. 
One way to handle this nature of the parametric
curves is to limit their domain of applicability to those regions, where they correspond to functions.
These regions are limited by the points, where the derivative of the  space-time rapidity
with respect to the parameter of the curve vanishes.
Due to this reason, we discuss the
solutions only in a finite domain around mid-rapidity in this manuscript. 
From the criteria $\partial_H \eta_x(H)  = 0$ we get the following
lower and upper limits for the applicability of our solution:
\begin{align}
	\eta_{min}&=-\frac{\lambda}{\sqrt{\left(\lambda-1\right)\left(\kappa-\lambda\right)}}\textnormal{arctan}\left(\sqrt{\frac{\kappa-\lambda}{\kappa\lambda-\kappa}}\right)+\textnormal{atanh}\left(\frac{1}{\sqrt{\kappa}}\right),\label{eq:etamin}\\
	\eta_{max}&= \frac{\lambda}{\sqrt{\left(\lambda-1\right)\left(\kappa-\lambda\right)}}\textnormal{arctan}\left(\sqrt{\frac{\kappa-\lambda}{\kappa\lambda-\kappa}}\right)-\textnormal{atanh}\left(\frac{1}{\sqrt{\kappa}}\right),\label{eq:etamax}	
\end{align}
where $\eta_{max} = - \eta_{min}$.
Thus the domain of validity of  our solutions 
is limited to the $\eta_{min} < \eta_x < \eta_{max}$ interval.
However, the parametric curves are defined outside this interval, too. This suggests the possibility
of the extension of our solutions, with the help of shock-wave equations, to the forward and backward
rapidity regions.  Their discussion is straightforward, but goes beyond the scope
of the present manuscript.

The finiteness of the mid-rapidity region implies, that 
there is an upper bound for the modulus of the possible values of the fluid rapidities:
\begin{equation}
|\Omega| \le \frac{\lambda}{\sqrt{\lambda-1}\sqrt{\kappa-\lambda}}\textnormal{arctan}\left(\sqrt{\frac{\kappa-\lambda}{\lambda-1}}
	\tanh\left(|\eta_{max}|\right) \right),
			 \label{e:Omega-range}
\end{equation}
which implies that this solution is finite in terms of the fluid rapidity $\Omega$.
This also implies that the solution yields a finite observable rapidity and pseudorapidity distribution.

Let us also clarify that our solutions are in fact not self-similar: 
although they contain factors that depend on the space-time
rapidity $\eta_x$ only through the scaling variable $s$, the solutions for the temperature and the 
entropy density contain additional factors that depend on $\eta_x$ through the parameter $H $ as well. 
This implies that our solutions are approximately self-similar only in a small domain  near mid-rapidity, 
but their self-similarity is explicitely violated by terms that become more and more important
with increasing deviations from mid-rapidity.

Let us note that in the $c_s^2=1/\kappa=1$ case, the solution described by eqs.
(\ref{e:sH}-\ref{e:etaH}) reproduces the 1+1 dimensional CNC solution of
refs.  ~\cite{Csorgo:2006ax,Nagy:2007xn}, as expected.  If $\kappa=1$ or if
$H\ll 1$, the fluid rapidity is an explicit function of the coordinate
rapidity, given as 

\begin{equation}
	\Omega\simeq \lambda\eta, \qquad \mbox{\rm for} \quad \eta_x \ll 1/(\lambda -1).
\end{equation}

However, let us also emphasize at this point that the $\kappa \rightarrow 1$ and the $\lambda
\rightarrow 1$ limits are not interchangeable: 
The CNC limit corresponds to the $\kappa \rightarrow 1$ limit of the extension
of the above solutions to the $1 \le \kappa < \lambda$ domain of the model
parameters, however the detailed description of this class of solutions is
beyond the scope of the present manuscript.  
In the $\kappa = 1$  limit, we recover the CNC, while in the $\lambda = 1$
limit, we recover the CGHK family of solutions. 

Let us also clarify, that our new family of solutions includes
the Hwa-Bjorken asymptotic solution as well. To see this,
one has to proceed carefully,
as the $H \rightarrow 0$ and the $\lambda \rightarrow 1$ limits 
of our new family of solutions are not interchangeable.
The Hwa-Bjorken solution is $\Omega = \eta_x$, and it corresponds to the
$H = 0 $ and $\lambda \rightarrow 1$ from above limiting case.
As $H \rightarrow 0$, $\eta_x \rightarrow \lambda \Omega$,
and if we take the $\lambda \rightarrow 1$ limit after the 
$H \rightarrow 0$ limit, we recover the flow field of the Hwa-Bjorken solution.
In this  case, the domain
of validity, the interval $(\eta_{min}\le \eta_x \le \eta_{max})$ also
approaches the whole space-time rapidity region, $(-\infty < \eta_x <\infty)$.
After these steps, we can consider the $\mathcal{T}(s)= 1$ special case,
to demonstrate that our new family of  solutions includes
the Hwa-Bjorken solution as a special limiting case. 
However, if one takes the $\lambda \rightarrow 1$ limit  
too early, before the $H  \rightarrow 0 $ limit, 
one could find an unphysical, divergent hence uninteresting limiting case.

\subsection{Graphical illustrations}
\begin{figure}[H] 
	\includegraphics[scale=0.15]{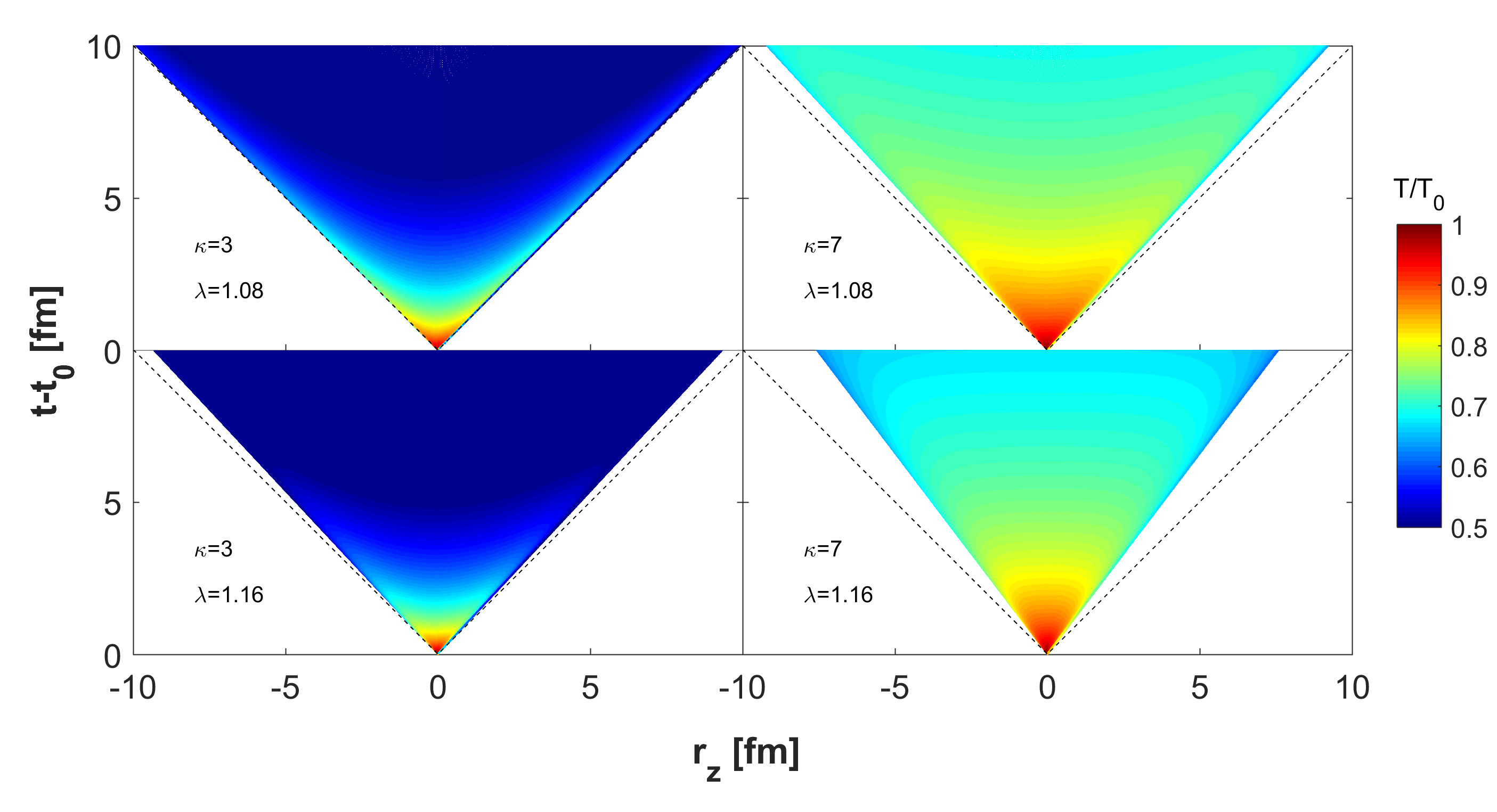} 
	\centering 
	\caption{
	Temperature maps in the forward light cone from our new, 
	longitudinally finite solutions for  $\kappa = \varepsilon/p = 3$ (left column)
	and for $\kappa = 7$ (right column) evaluated for the acceleration parameters $\lambda = 1.08$ (top row) 
	corresponding to a broader rapidity distribution and for $\lambda = 1.16$ (bottom row) corresponding to
	a narrower rapidity distribution, approximately corresponding to heavy ion collisions at $\sqrt{s_{NN}} = 200$
	GeV RHIC and $\sqrt{s_{NN}} = 2.76 $ TeV LHC energies.
	}
\label{fig:T-t-rz} 
\end{figure}

\begin{figure}[H] 
	\includegraphics[scale=1.1]{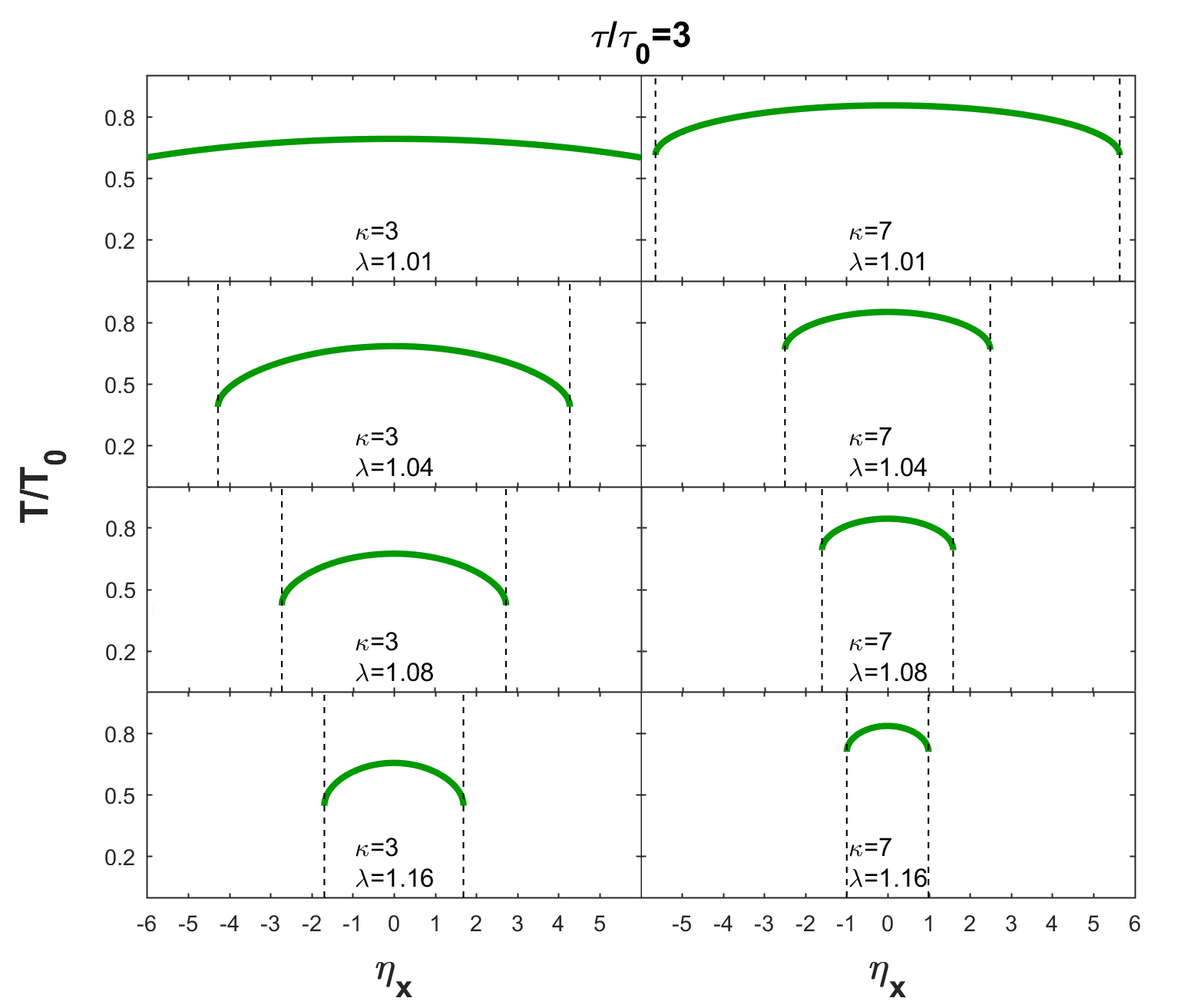} 
	\centering 
	\caption{
	Space-time rapidity $\eta_x$ dependence of the temperature profile at a fixed $\tau/\tau_0 = 3$ fm/c,
	from our new, longitudinally finite solutions of relativistic hydrodynamics.
	These plots highlight the finiteness of these solutions, with parameters the same as
	for Figure~\ref{fig:T-t-rz}: $\kappa = \varepsilon/p = 3$ in the left column,
	$\kappa = 7$ in the right column,  evaluated for the acceleration parameters $\lambda = 1.08$  in the top row 
	and for $\lambda = 1.16$ in the bottom row. 
	}
\label{fig:T_eta} 
\end{figure}

The temperature map of our new family of solutions is indicated on Figure ~\ref{fig:T-t-rz}. The finiteness
of these solutions means that they are defined in a cone,
$(\eta_{min}\le \eta_x \le \eta_{max})$ that lies within the forward light cone. The pseudorapidity dependence of the temperature
at various constant values of the proper-time is indicated on Figure ~\ref{fig:T_eta}.
It is remarkable, how similar these exact solutions are to the figures of those Monte-Carlo simulations,
that mapped for the space-time picture of heavy ion collisions, see for one of the  first examples
the figures of refs. ~\cite{Csorgo:1988pu,Csorgo:1989kq}.

\begin{figure}[H] 
	\includegraphics[scale=0.15]{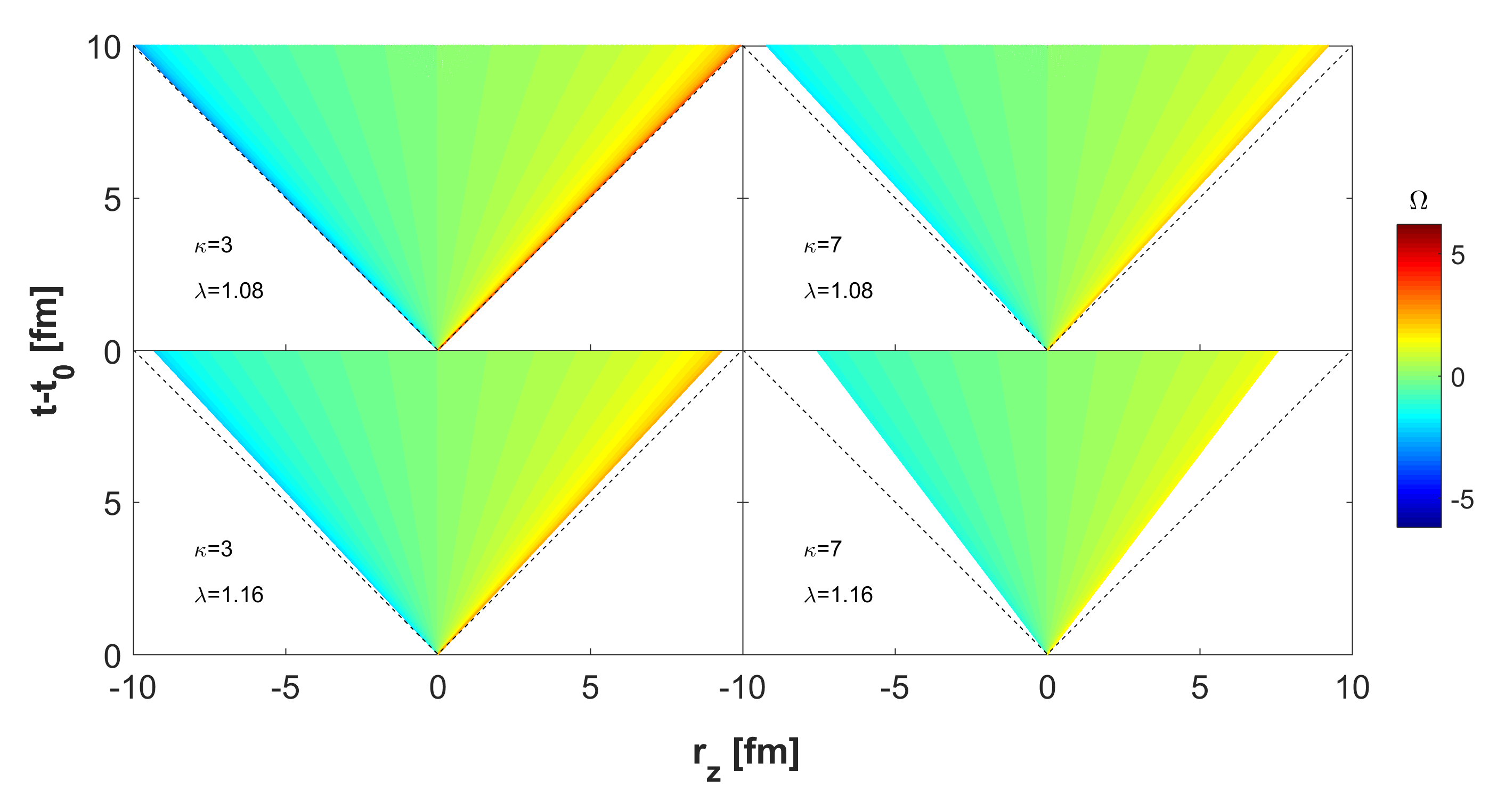} 
	\centering 
	\caption{
	Fluid rapidity $\Omega$ maps in the forward light cone from our new, 
	longitudinally finite solutions are shown for for  $\kappa = \varepsilon/p = 3$ (left column)
	and for $\kappa = 7$ (right column) evaluated for the acceleration parameters 
	$\lambda = $ 1.01, 1.02, 1.04 and 1.08  (from top to bottom rows) 
	corresponding to nearly flat and with increasing $\lambda$, gradually narrowing rapidity distributions.
	}
\label{fig:Omega-t-rz} 
\end{figure}

\begin{figure}[H] 
	\includegraphics[scale=1.1]{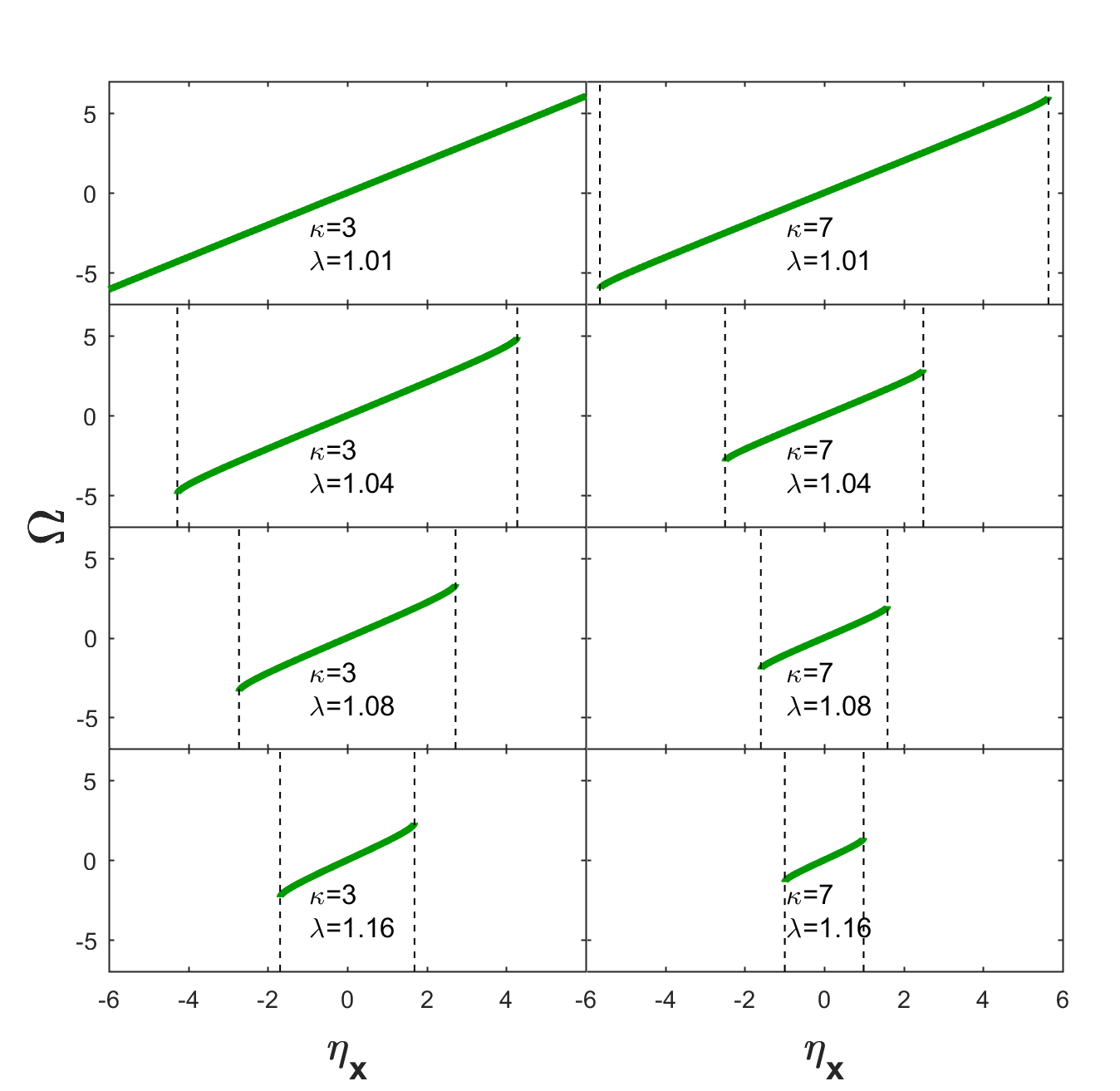} 
	\centering 
	\caption{
	Fluid rapidity distributions versus space-time rapidity $\eta_x$
	from the new, longitudinally finite solutions for  $\kappa =  3$ (left column)
	and for $\kappa = 7$ (right column) evaluated for the acceleration parameters 
	$\lambda = $ 1.01, 1.02, 1.04 and 1.08  (from top to bottom rows) 
	corresponding to gradually  narrowing rapidity distributions with increasing acceleration parameter. 
	Note that in our solutions, the $\Omega(\eta_x)$ functions are approximately, but not exactly, linear functions 
	of the space-time rapidity $\eta_x$, but they are exactly independent of the proper-time $\tau$.
	}
\label{fig:Omega_eta} 
\end{figure}

The map of the fluid-rapidity distribution  of  our new family of solutions,
$\Omega(t,rz)$ is illustrated on Figure ~\ref{fig:Omega-t-rz}.  This plot shows not
only the finiteness of these solutions, that corresponds to a cone within the forward
light cone, but also that in our case the fluid rapidity is a function of the space-time 
rapidity only, so that $ \Omega = \Omega(\eta_x)$,
independently of the proper-time $\tau$.  The pseudorapidity dependence of the fluid-rapidity at
various constant values of the proper-time is indicated on Figure
~\ref{fig:Omega_eta}. This figure also indicates that our new exact solutions,
altough formally different, numerically are
very close to the $\Omega = \lambda\eta_x$ flow field, that
is to the CNC solutions of refs.  ~\cite{Csorgo:2006ax,Nagy:2007xn}.

In these earlier CNC solutions, the $\Omega = \lambda \eta_x$, flow rapidity distribution was an exact solution,
but only for a super-hard,  hence unrealistic equation of state, corresponding to the $\kappa = 1$, 
or speed of sound $c_s = 1$ equation of state.  
Although this equation of state has been changed drastically,
the speed of sound was decreased from speed of light to any temperature independent value, 
the flow field changed surprisingly little: most of its  change is only at forward and backward rapidities 
and it is only a few \% for the experimentally relevant values of the parameters $\lambda$ and $\kappa$.

As the acceleration parameter $\lambda \rightarrow 1$, the
space-time rapidity region opens up, as it approaches the whole horizontal axes, at the same time the temperature
approaches a constant value that becomes independent of the value of the space-time rapidity coordinate $\eta_x$:
we recover the boost-invariant Hwa-Bjorken solution.
This is a common property of Figures ~\ref{fig:T-t-rz} - \ref{fig:Omega_eta} .

\begin{figure}[H] 
	\includegraphics[scale=0.10]{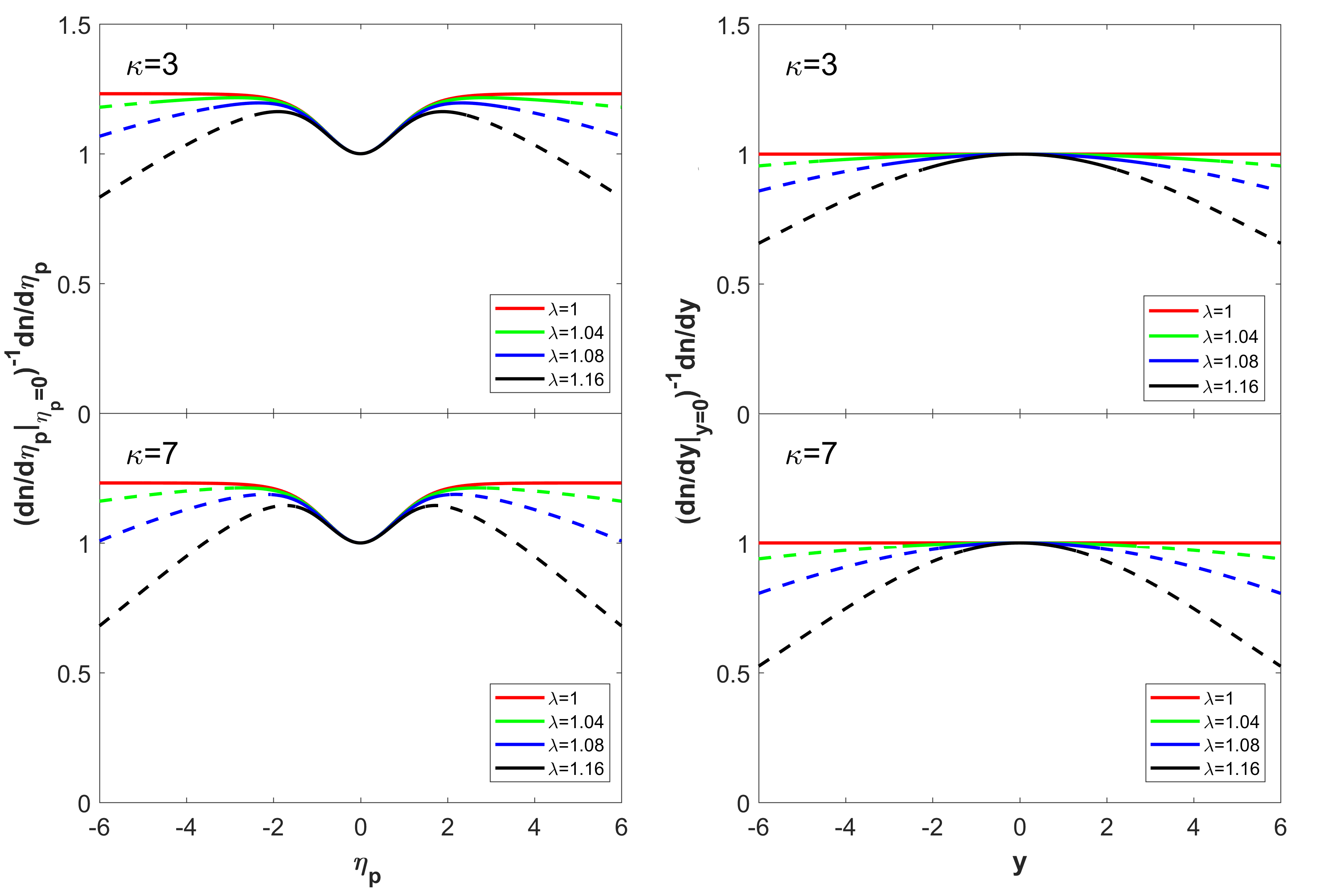} 
	\centering 
	\caption{
	Left panel indicates the pseudorapidity $\eta_p$ densities while the right panel
	the rapidity $y$ densities, as obtained from our  
	new class of exact solutions of relativistic hydrodynamics,
	for  $\kappa =  3$ (top panels)
	and for $\kappa = 7$ (bottom panels), evaluated for the acceleration parameters $\lambda = $ 1.0, 1.04, 1.08 and 1.16 
	(curves from top to bottom) corresponding to a  rapidity distribution that is flat in the boost-invariant limit,
	and that becomes gradually narrowing with increasing acceleration parameter $\lambda$.
	}
\label{fig:dndeta-dndy} 
\end{figure} 

\section{Observables: rapidity  and pseudorapidity distributions}

In order to be able to compare our solution to data, let us calculate
analytically the observable quantities, that correspond to the rapidity and to
the pseudorapidity distributions, both defined in momentum space.  The
detailed calculation is not included in this manuscript, but the main steps are
summarized as follows. We assume the simplest possible form of the temperature
scaling function, using $\mathcal{T}(s) = 1$. We utilize a Boltzmann approximation to evaluate the
phase-space densities on the freeze-out hypersurface,  which is assumed to be
pseudo-orthogonal to the four-velocity $u^{\mu}$.  
Consequently, the normal vector of the freeze-out hypersurface is parallel to 
$u^{\mu}$, i.e. $d\sigma^{\mu}=u^{\mu}d\eta_x$, where $d\sigma^{\mu}$ is the infinitesimal form of
the normal vector of the freeze-out hypersurface:
\begin{equation}
d\sigma^{\mu}=\frac{1}{A(\eta_x)}\left(\partial_{\eta_x}r_z,\:\partial_{\eta_x}t\right)d\eta_x,
\end{equation}
where $A(\eta_x)$ is a normalization factor. We find that the equation of the freeze-out hypersurface
is the following:
\begin{equation}
\frac{\tau(H)}{\tau_f}=\cosh^{\frac{\kappa}{\lambda-\kappa}}\left(H\right)\left[1+\frac{\kappa-1}{\lambda-1}\sinh^2(H)\right]^{\frac{\lambda}{2(\kappa-\lambda)}}.
\end{equation}
Here $\tau_f$ stands  for the proper-time of the kinetic freeze-out at
midrapidity, corresponding to the $H=\eta_x=\Omega= 0$ parameter values.  
During the calculation of the rapidity
distribution, we assume that we are not too far away from the midrapidity region,
and with the help of a saddle-point approximation we perform the integral over
$\eta_x$ to obtain the following formula:
\begin{equation}\label{eq:rapidity-dist}
	\frac{dn}{dy}\approx\left.\frac{dn}{dy}\right|_{y=0} \cosh^{-\frac{1}{2}\alpha(\kappa)-1}\left(\frac{y}{\alpha(1)}\right)\exp\left(-\frac{m}{T_f} \left[\cosh^{\alpha(\kappa)}\left(\frac{y}{\alpha(1)}\right)-1\right]\right),
\end{equation}
where $y$ is the rapidity, and $\alpha(\kappa)$ stands for
\begin{equation}\label{e:alpha-kappa}
	\alpha(\kappa) = \frac{2\lambda-\kappa}{\lambda-\kappa}.
\end{equation}
A similar, but $\kappa$ independent exponent, $\alpha(\kappa=1)\equiv \alpha(1) $ was denoted with $\alpha$
in the CNC solutions of refs.~\cite{Csorgo:2006ax,Nagy:2007xn}.

The normalization factor of this distribution is
\begin{equation}
\left.\frac{dn}{dy}\right|_{y=0}=\frac{R^2\pi\tau_f}{(2\pi\hbar)^3}\sqrt{\frac{(2\pi T_f m)^3}{\lambda(2\lambda-1)}}\exp\left(-\frac{m}{T_f}\right),
\end{equation}
where the $R^2\pi$ multiplication factor is the transverse area and $m$ is the
mass of the observed particle (predominantly pion). 

We emphasize, that the result of
eq.~\eqref{eq:rapidity-dist} reproduces the flat Hwa-Bjorken rapidity plateaux 
~\cite{Bjorken:1982qr,Hwa:1974gn} in the $\lambda\rightarrow 1$ from above  limit.
In the $\kappa \rightarrow 1$ limit, our calculations also reproduce
the rapidity distribution of the CNC solution of refs.~(\cite{Csorgo:2006ax} , \cite{Nagy:2007xn} ).
Using  a mean value theorem for definite integrals and 
a saddle point approximation, the pseudorapidity distribution was found to be
proportional to eq.~\eqref{eq:rapidity-dist}, and the constant of proportionality 
to be the average value of the  $\mathcal{J}=\frac{dy}{d\eta_p}$ Jacobi determinant.
In this way the pseudorapidity distribution can be expressed in the following
form:
\begin{equation}
	\frac{dn}{d\eta_p}\approx\left.\frac{dn}{dy}\right|_{y=0} 
	\frac{\langle p_T(y)\rangle\cosh\left(\eta_p\right)}{\sqrt{m^2+\langle p_T(y)\rangle^2 \cosh(\eta_p)}} 
		\cosh^{-\frac{1}{2}\alpha(\kappa)-1}\left(\frac{y}{\alpha(1)}\right)
		\exp\left(-\frac{m}{T_f} \left[\cosh^{\alpha(\kappa)}\left(\frac{y}{\alpha(1)}\right)-1\right]\right), 	\label{e:dndeta}
\end{equation}
where $\langle p_T(y) \rangle$ is the rapidity dependent average transverse momentum:
\begin{equation}
	\langle p_T (y) \rangle \approx \frac{\sqrt{T_f^2 + 2mT_f}}{\displaystyle\strut 1+\frac{\displaystyle\strut \alpha(\kappa)}{\displaystyle\strut 2\alpha(1)^2}\frac{\displaystyle\strut T_f+m}{\displaystyle\strut T_f+2m}y^2}.
	\label{e:meanpt_y}
\end{equation}
Note, that the same functional form, a Lorentzian shape was assumed for the rapidity dependence of the slope
of the transverse  momentum spectrum in the Buda-Lund hydro model of ref.
~\cite{Csorgo:1995bi}. 
This form was obtained based on arguments of the
analyticity of the inverse temperature profile in rapidity 
and the coefficient of the $y^2$ dependence was considered 
also very recently in refs.~\cite{Csanad:2016add,Ze-Fang:2017ppe} 
as a free fit parameter. Our exact solutions of fireball hydrodynamics 
express this coefficient with two observables, the
freeze-out temperature $T_f$ and the mass,  and in addition, with the speed of sound parameter of the solution, $\kappa = 1/c_s^2$.
Thus from the measurement of the rapidity dependence of the transverse momentum spectra, the equation of state parameter
can in fact be fitted to data, at least in principle. 
Note that such an approximately Lorentzian rapidity dependence of the slope of the transverse momentum spectra
has also been experimentally tested in h+p reactions at $\sqrt{s} = 22 $ GeV by the EHS/NA22 collaboration in ref. ~\cite{Agababyan:1997wd},
so apparently this feature of our exact solutions is realistic: it has experimental support.

The normalized rapidity and pseudorapidity distributions are illustrated on Fig.~\ref{fig:dndeta-dndy}.
They look rather realistic and work has just been started to fit them to experimental data in
proton-proton and heavy ion collisions at BNL RHIC and at CERN LHC energies. Some of the first results
are shown in the next section.

\section{First comparisons to p+p data at RHIC and LHC}
\begin{figure}[H] 
	\includegraphics[scale=0.3]{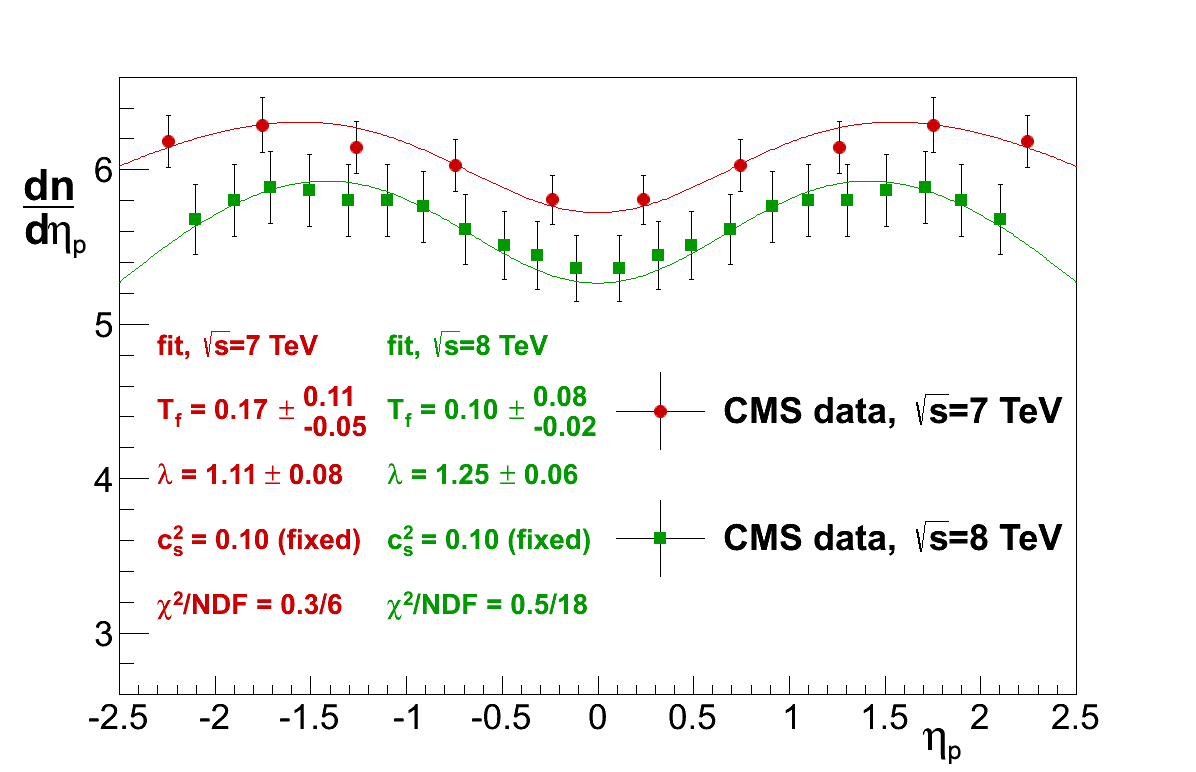} 
	\centering 
	\caption{
		CMS data on $dn/d\eta_p$ pseudorapidity densities,
		as measured in $\sqrt{s} = $$7$ ~\cite{Khachatryan:2010us}
		and $8$ TeV ~\cite{Chatrchyan:2014qka}
		p+p collisions at LHC are well described with eqs. (\ref{e:dndeta},\ref{e:meanpt_y}), as obtained from our
		exact family of solutions, corresponding to eqs.~(\ref{e:etaH},\ref{e:sH}),
		using the equation of state $\varepsilon = \kappa p$ 
		using a realistic $\kappa = 10 p$ value, that  is in agreement with $c_s^2 = 1/\kappa ~ 0.1$, 
		according to PHENIX measurements
		of the average value of the speed of sound in high energy Cu+Cu and  Au+Au collisions
		at $\sqrt{s_{NN}}= 200 $ GeV ~\cite{Adare:2006ti}.
		}
\label{fig:dndeta-dndy-kappa10} 
\end{figure}

\begin{figure}[H] 
	\includegraphics[scale=0.3]{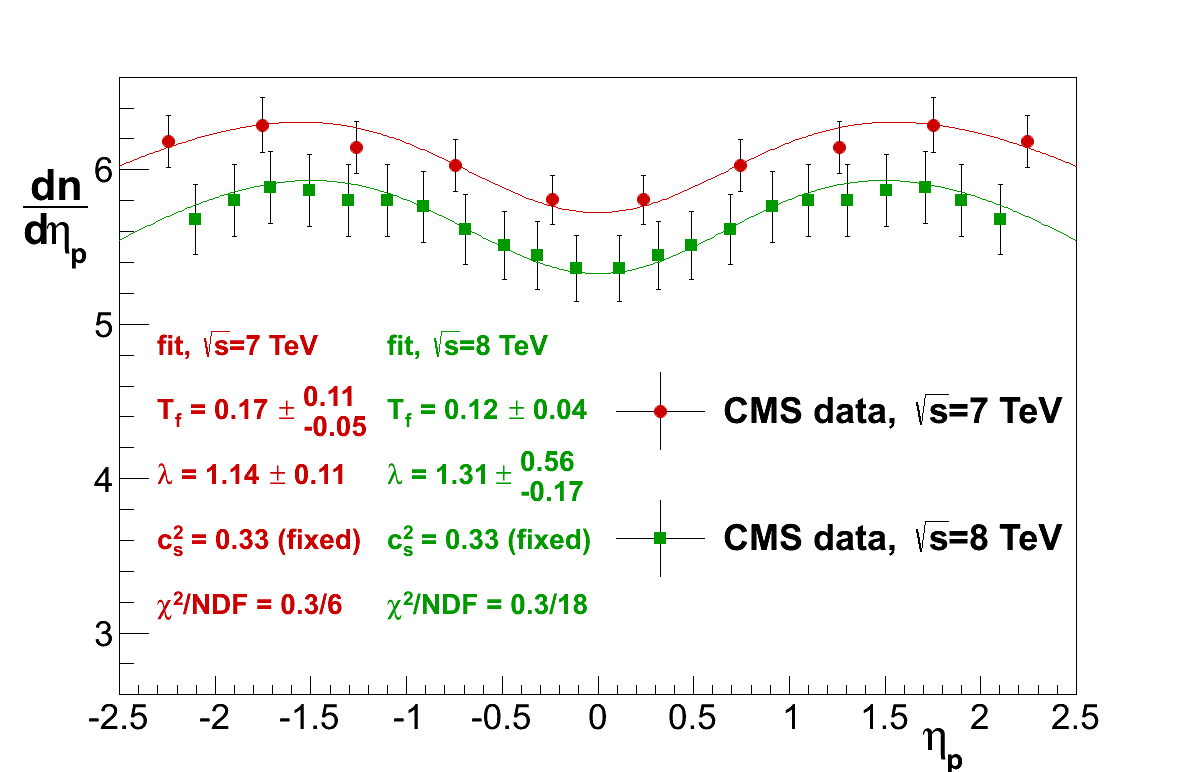} 
	\centering 
	\caption{
		CMS data on $dn/d\eta_p$ pseudorapidity densities,
		as measured in $\sqrt{s} = $$7$ 
		and $8$ TeV p+p collisions at LHC~\cite{Khachatryan:2010us,Chatrchyan:2014qka} 
		are  well described with eqs. (\ref{e:dndeta},\ref{e:meanpt_y}).
		These formulas  are obtained from our new exact family of solutions
		of relativistic fireball hydrodynamics, eqs.~(\ref{e:etaH},\ref{e:sH}),
		and the equation of state $\varepsilon = 3 p$, corresponding
		to the idealized case of a massless relativistic with a speed of sound of $c_s^2 = 1/3$.
		}
\label{fig:dndeta-dndy-kappa3} 
\end{figure}

The first comparisons of our new, exact solutions of relativistic hydrodynamics to experimental 
data are shown on Figures ~\ref{fig:dndeta-dndy-kappa10} and ~\ref{fig:dndeta-dndy-kappa3}.

Apparently, CMS data on $dn/d\eta_p$ pseudorapidity densities,
as measured in $\sqrt{s} = $$7$ ~\cite{Khachatryan:2010us} and $8$ TeV ~\cite{Chatrchyan:2014qka}
p+p collisions at LHC are described with eqs. (\ref{e:dndeta},\ref{e:meanpt_y}), as obtained from our
exact family of solutions, corresponding to eqs.~(\ref{e:etaH},\ref{e:sH}). 
Let us also clarify, that the two investigated CMS datasets  were measured with somewhat different trigger conditions
at $\sqrt{s} = 7 $ and $8$ TeV, which explains the difference between their absolute normalizations.

The confidence level of these fits, CL > 0.1~\% indicates that our new solution is not inconsistent with these CMS data, 
and the parameters can be interpreted as meaningful ones.
However, given the large errors of the measurement, 
the parameters of the $\sqrt{s} = 7 $ and $\sqrt{s} = 8$ TeV reactions cannot be distinguished based
on data fitting, given that their values are within three standard deviations, and 
within the systematic error of this analysis, the same.

One should also be careful to draw too strong conclusions here because our solutions are limited in
rapidity range. The domain of the validity of our solution is given by eqs. (\ref{eq:etamin},\ref{eq:etamax})
and depending on the values of the fit parameters $\kappa $ and $\lambda$ these range from 
$|\eta_x| < 1.03 $ to $|\eta_x|< 2.5$. Given that our solutions are finite, the propagation of shock-waves
should be investigated in the target and in the projectile rapidity region, but instead we limit the range
and plot the comparison to data only in the limited $|\eta_p | < 2.5$ pseudorapidity interval.

Let us also note that we can describe the data with physically very different
equations of state. For example, using the equation of state $\varepsilon = \kappa p$ 
with a realistic, $\kappa = 10 $ value, in agreement with $c_s^2 = 1/\kappa  \approx 0.1$ 
of ref.~\cite{Adare:2006ti}, we get as good a description in Fig.~\ref{fig:dndeta-dndy-kappa10} as with the ideal, massless relativistic gas equation of state,
corresponding to $\kappa = 3$ on Fig.~\ref{fig:dndeta-dndy-kappa3}.
Similarly good quality fits were obtained by members of our group to describe the same
pseudorapidity distribution, but measured in the complete pseudorapidity range,
using the unrealistic, superhard equation of state $\kappa = 1$ ~\cite{Csanad:2016add}. 
So one of the apparent conclusions is that 
the pseudorapidity distributions are not very sensitive, at the present level
of their precision, to the equation of state parameter $\kappa$ as long
as the flow field remains approximately linear, $\Omega \approx \lambda \eta_x$.


\section{Summary}
We have found a new family of analytic and accelerating, exact and finite solutions of perfect fluid hydrodynamics
for $1$$+$$1$ dimensionally expanding fireballs. These solutions are defined in a cone within the forward lightcone,
and can be used to fit rapidity or pseudorapidity densities at RHIC and at LHC. Near mid-rapidity, the formulas
for these observables approximate well those calculations obtained from the earlier the CNC solution,
and near mid-rapidity, the longitudinal flow approximates the $v = \tanh(\lambda \eta_x)$ CNC flow profile.
Due to small correction terms, our solutions in principle allow to determine both the acceleration parameter $\lambda $
and the equation of state parameter $\kappa$ directly  from fitting the measured (pseudo)rapidity densities 
as well as the rapidity dependence of the mean transverse momentum high energy
proton-proton and heavy ion collisions at RHIC and at LHC.
The Hwa-Bjorken, the CGHK and the CNC solutions are recovered in appropriate limits of the model parameters.

Further generalizations of these solutions to 1$+ $$3$ dimensional, transversally expanding and/or rotating exact solutions,
as well as to solutions with a temperature dependent speed of sound are desirable 
and are being explored at the time of closing this manuscript.

\section*{Acknowledgments}
Our research has been partially supported by the 
bilateral Chinese–Hungarian intergovernmental grant No.T\'ET 12CN-1-2012-0016, 
the CCNU PhD Fund 2016YBZZ100 of China,
the COST Action CA15213, THOR Project of the European Union,
the Hungarian NKIFH grants No. FK-123842 and FK-123959,
the Hungarian EFOP 3.6.1-16-2016-00001 project,
the NNSF of China under grant No.11435004 
and by the exchange programme of the Hungarian and the Ukrainian Academies of Sciences, grants
NKM-82/2016 and NKM-92/2017.
M. Csan\'ad was partially supported by the J\'anos Bolyai Research Scholarship and the \'UNKP-17-4 New National Excellence
Program of the Hungarian Ministry of Human Capacities.

\vfill
\end{document}